\documentclass[pdflatex,sn-nature]{sn-jnl}

\usepackage{graphicx}
\graphicspath{{../}}
\usepackage{multirow}
\usepackage{amsmath,amssymb,amsfonts}
\usepackage{amsthm}
\usepackage{mathrsfs}
\usepackage[title]{appendix}
\usepackage{xcolor}
\usepackage{textcomp}
\usepackage{manyfoot}
\usepackage{booktabs}
\usepackage{algorithm}
\usepackage{algorithmicx}
\usepackage{algpseudocode}
\usepackage{listings}
\usepackage{subcaption}
\usepackage{xspace}
\usepackage{afterpage}

\theoremstyle{thmstyleone}

\theoremstyle{thmstyletwo}

\theoremstyle{thmstylethree}

\newcommand{\kce}{SN~2024kce\xspace}
\newcommand{\ion}[2]{#1~\MakeUppercase{\romannumeral#2}}
\newcommand{\fion}[2]{[#1~\MakeUppercase{\romannumeral#2}]}
\newcommand{\ionl}[3]{#1~\MakeUppercase{\romannumeral#2}~$\lambda$#3}

\DeclareRobustCommand{\okina}{%
  \raisebox{\dimexpr\fontcharht\font`f-\height}{%
    \scalebox{0.8}{`}%
  }%
}

\newcommand{\purdue}{1}
\newcommand{\uva}{2}
\newcommand{\hawaii}{3}
\newcommand{\msu}{4}
\newcommand{\davis}{5}
\newcommand{\lco}{6}
\newcommand{\rutgers}{7}
\newcommand{\melbourne}{8}
\newcommand{\carnegie}{9}
\newcommand{\uiuc}{10}
\newcommand{\AI}{11}
\newcommand{\florida}{12}
\newcommand{\anu}{13}
\newcommand{\anupa}{14}
\newcommand{\osu}{15}

\begin{document}

\title[Helium in a thermonuclear supernova]{Early panchromatic spectra reveal helium in the thermonuclear supernova 2024kce}

\author*[\purdue]{\fnm{Abigail} \sur{Polin}\orcid{https://orcid.org/0000-0002-1633-6495}}\email{abigail@purdue.edu}
\author[\purdue]{\fnm{Dillon}\sur{Hasenour}\orcid{https://orcid.org/0009-0003-6165-0461}}
\author[\uva]{\fnm{Sahana}\sur{Kumar}\orcid{https://orcid.org/0000-0001-8367-7591}}
\author[\hawaii]{\fnm{Willem}\sur{Hoogendam}\orcid{https://orcid.org/0000-0003-3953-9532}}
\author[\msu]{\fnm{Jing}\sur{Lu}\orcid{https://orcid.org/0000-0002-3900-1452}}
\author[\davis]{\fnm{Daniel}\sur{Polin}\orcid{https://orcid.org/0000-0001-7445-4724}}
\author[\purdue]{\fnm{Danielle}\sur{Dickinson}\orcid{https://orcid.org/0000-0003-0913-4120}}
\author[\purdue]{\fnm{Braden}\sur{Garretson}\orcid{https://orcid.org/0000-0001-6922-8319}}
\author[\lco]{\fnm{Nidia}\sur{Morrell}\orcid{https://orcid.org/0000-0003-2535-3091}}
\author[\rutgers]{\fnm{Huei}\sur{Sears}\orcid{https://orcid.org/0000-0001-8023-4912}}
\author[\melbourne]{\fnm{Katie} \sur{Auchettl}\orcid{https://orcid.org/0000-0002-4449-9152}}
\author[\purdue]{\fnm{Teresa} \sur{Boland}\orcid{https://orcid.org/0000-0002-1250-4690
}}
\author[\carnegie]{\fnm{Christopher} \sur{Burns}\orcid{https://orcid.org/0000-0003-4625-6629}} 
\author[\hawaii]{\fnm{Kenneth C.} \sur{Chambers}\orcid{https://orcid.org/0000-0001-6965-7789}}
\author[\hawaii]{\fnm{Dhvanil} \sur{Desai}\orcid{https://orcid.org/0000-0002-2164-859X}}
\author[\uiuc,\AI]{\fnm{Jason T.}\sur{Hinkle}\orcid{https://orcid.org/0000-0001-9668-2920
}}
\author[\purdue]{\fnm{Thressay}\sur{Hoover}\orcid{https://orcid.org/0000-0003-4114-1240}}
\author[\florida]{\fnm{Eric} \sur{Hsiao}\orcid{https://orcid.org/0000-0003-1039-2928}}
\author[\hawaii]{\fnm{Mark E.} \sur{Huber}\orcid{https://orcid.org/0000-0003-1059-9603}}
\author[\hawaii]{\fnm{David~O.} \sur{Jones}\orcid{https://orcid.org/0000-0002-6230-0151}}
\author[\purdue]{\fnm{Colin}\sur{Macrie}\orcid{https://orcid.org/0000-0002-9209-2787}}
\author[\anu]{\fnm{Bailey} \sur{Martin}\orcid{https://orcid.org/0009-0006-4963-3206}}
\author[\hawaii]{\fnm{Cameron} \sur{Pfeffer}\orcid{https://orcid.org/0000-0002-7305-8321}}
\author[\purdue]{\fnm{Miranda} \sur{Pikus}\orcid{https://orcid.org/0000-0001-7676-948X}}
\author[\hawaii]{\fnm{Benjamin~J.} \sur{Shappee}\orcid{https://orcid.org/0000-0003-4631-1149}}
\author[\purdue]{\fnm{Madeline}\sur{Taylor}}
\author[\anu, \anupa]{\fnm{Brad E.} \sur{Tucker}\orcid{https://orcid.org/0000-0002-4283-5159}}
\author[\osu]{\fnm{Michael} \sur{Tucker}\orcid{https://orcid.org/0000-0002-2471-8442
}} 
\author[\florida]{\fnm{Huangfei} \sur{Xiao}\orcid{https://orcid.org/0009-0006-9436-7197}}

\affil[\purdue]{
  \orgdiv{Department of Physics and Astronomy},
  \orgname{Purdue University},
  \orgaddress{
    \street{525 Northwestern Avenue},
    \city{West Lafayette},
    \state{IN},
    \postcode{47906},
    \country{USA}
  }
}
\affil[\uva]{
  \orgdiv{Department of Astronomy},
  \orgname{University of Virginia},
  \orgaddress{
    \street{530 McCormick Rd.},
    \city{Charlottesville},
    \state{VA},
    \postcode{22904},
    \country{USA}
  }
}

\affil[\hawaii]{
  \orgdiv{Institute for Astronomy},
  \orgname{University of Hawai\okina i},
  \orgaddress{
    \street{2680 Woodlawn Drive},
    \city{Honolulu},
    \state{HI},
    \postcode{96822},
    \country{USA}
  }
}
\affil[\msu]{
  \orgdiv{Department of Physics and Astronomy},
  \orgname{Michigan State University},
  \orgaddress{
         \street{567 Wilson Road},
    \city{East Lansing},
    \state{Michigan},
    \postcode{48824},
    \country{USA}
  }
}

\affil[\davis]{
  \orgdiv{Department of Physics and Astronomy},
  \orgname{{University of California, Davis}},
  \orgaddress{
    \street{1 Shields Avenue},
    \city{Davis},
    \state{CA},
    \postcode{95616},
    \country{USA}
  }
}

\affil[\lco]{  \orgdiv{Las Campanas Observatory},
 \orgname{Carnegie Observatories},  
\orgaddress{    
\street{Casilla 601},    
\city{La Serena},      
\country{Chile}  } }

\affil[\rutgers]{
  \orgdiv{Department of Physics and Astronomy},
  \orgname{Rutgers, the State University of New Jersey},
  \orgaddress{
    \street{136 Frelinghuysen Road},
    \city{Piscataway},
    \state{NJ},
    \postcode{08854},
    \country{USA}
  }
}

\affil[\melbourne]{  \orgdiv{School of Physics},
 \orgname{University of Melbourne},  
\orgaddress{    
\street{Grattan Street},    
\city{Parkville},  
 \state{Victoria},  
 \postcode{3010},    
\country{Australia}  } }

\affil[\carnegie]{  \orgdiv{Observatories},  \orgname{Carnegie Science},  \orgaddress{    \street{813 Santa Barbara Street},    \city{Pasadena},    \state{CA},    \postcode{91104},    \country{USA}  } }

\affil[\uiuc]{  \orgdiv{Department of Astronomy},  \orgname{University of Illinois Urbana-Champaign},  \orgaddress{    \street{1002 West Green Street},    \city{Urbana},    \state{IL},    \postcode{61801},    \country{USA}  } }

\affil[\AI]{ \orgname{NSF-Simons AI Institute for the Sky (SkAI)},  \orgaddress{172 E. Chestnut St.},    \city{Chicago},    \state{IL},    \postcode{60611},    \country{USA}  } 

\affil[\florida]{
  \orgdiv{Department of Physics},
  \orgname{Florida State University},
  \orgaddress{
    \street{77 Chieftan Way},
    \city{Tallahassee},
    \state{FL},
    \postcode{32306},
    \country{USA}
  }
}

\affil[\anu]{  \orgdiv{Research School of Astronomy and Astrophysics},  \orgname{The Australian National University},  \orgaddress{    \street{Mount Stromlo Road},    \city{Canberra},    \state{ACT},    \postcode{2611},    \country{Australia}  } }

\affil[\anupa]{
  \orgdiv{Australian National Centre for the Public Awareness of Science},
  \orgname{Australian National University},
  \orgaddress{
    \street{Building 42A},
    \city{Acton},
    \state{ACT},
    \postcode{2601},
    \country{Australia}
  }
}

\affil[\osu]{  
\orgdiv{Center for Cosmology and AstroParticle Physics (CCAPP)},  
\orgname{The Ohio State University},  
\orgaddress{    
\street{191 West Woodruff Avenue},    
\city{Columbus},    
\state{OH},    
\postcode{43210},    
\country{USA}  
} 
}

\abstract{Faint Type I supernovae blur the boundaries between spectroscopic classes, obscuring their progenitors and explosion mechanisms. One proposed origin is a double detonation, in which ignition of an accreted helium shell triggers the explosion of a carbon oxygen white dwarf. Surviving helium is a direct prediction of this mechanism, but its strongest near-infrared feature overlaps with carbon and is ambiguous in isolation. Here we present \kce, observed with simultaneous optical and near-infrared spectroscopy eleven days before maximum light. We identify \ion{He}{1} absorption at three wavelengths, all consistent with a velocity of 16,400~km/s, well above the photosphere. This is the first detection of helium before maximum light in a thermonuclear supernova. Helium at these velocities requires nearby radioactive material for non-thermal excitation, pointing to ash from a detonated surface helium shell. A double detonation model reproduces the optical spectral evolution, suppressed blue flux, and \ionl{He}{1}{1.083}~$\mu$m absorption feature. \kce shares properties with both subluminous Type Ia supernovae and Ia-like calcium-rich transients, providing evidence that the same explosion mechanism operates across these classes. From early optical and near-infrared spectra to Rubin's First Look images, \kce showcases the full reach of modern observatories in exposing the physics of rare explosions.}

\keywords{Type Ia supernovae, white dwarfs, double detonation, calcium-rich transients}

\maketitle

\newpage 
\section*{Main:}
Type Ia supernovae (SNe) are the thermonuclear explosions of white dwarfs (WDs), and their role as cosmological distance indicators makes the question of how WDs explode, one of the most consequential open problems in astrophysics. A leading theory is the double detonation mechanism, in which the ignition of an accreted helium shell drives a shock into the WD and detonates its carbon oxygen (C/O) core \cite{Nomoto1982b, Woosley&Weaver94, Livne1990}. Operating across a range of core and shell masses, this mechanism has been proposed for normal SNe Ia \cite{Shen2018,2021ApJ...919..126B} as well as the faint outskirts of the thermonuclear population, from subluminous 91bg-like SNe Ia \cite{Polin2019,Shen14} to calcium-rich transients (Ca-RTs) \cite{Bildsten2007,2024A&A...683A..44M}. The mechanism carries a testable prediction: some helium should survive the shell detonation and, in systems with a large helium shell, imprint high velocity absorption on the earliest spectra \cite{Polin2019, 2023MNRAS.524.4447C}. 
Testing this prediction requires observations that are both early and panchromatic. The strongest NIR helium line, \ionl{He}{1}{1.083~$\mu$m}, can overlap with \ionl{C}{1}{1.069~$\mu$m}, making the feature ambiguous in isolation. A secure identification requires simultaneous optical and NIR spectroscopy that covers the weaker corroborating helium transitions while the outer ejecta still shape the spectrum.

Here we present \kce, a peculiar Type I SN observed with simultaneous optical and NIR spectroscopy eleven days before maximum light, when the spectrum still probes the outermost ejecta. We establish a thermonuclear origin for \kce and identify high velocity helium in its outer layers through absorption at three separate wavelengths, providing the first detection of helium before maximum light in a thermonuclear SN. A double detonation model reproduces the spectral evolution and exhibits \ionl{He}{1}{1.083}~$\mu$m absorption from surviving helium that is co-located with radioactive shell ashes. We therefore identify \kce as the product of a helium shell double detonation. Because \kce shares properties with both subluminous SNe Ia and Ia-like Ca-RTs, establishing its origin provides evidence that the same explosion mechanism operates across these observational classes.

\begin{figure}[p]
    \centering
    \includegraphics[width=\textwidth]{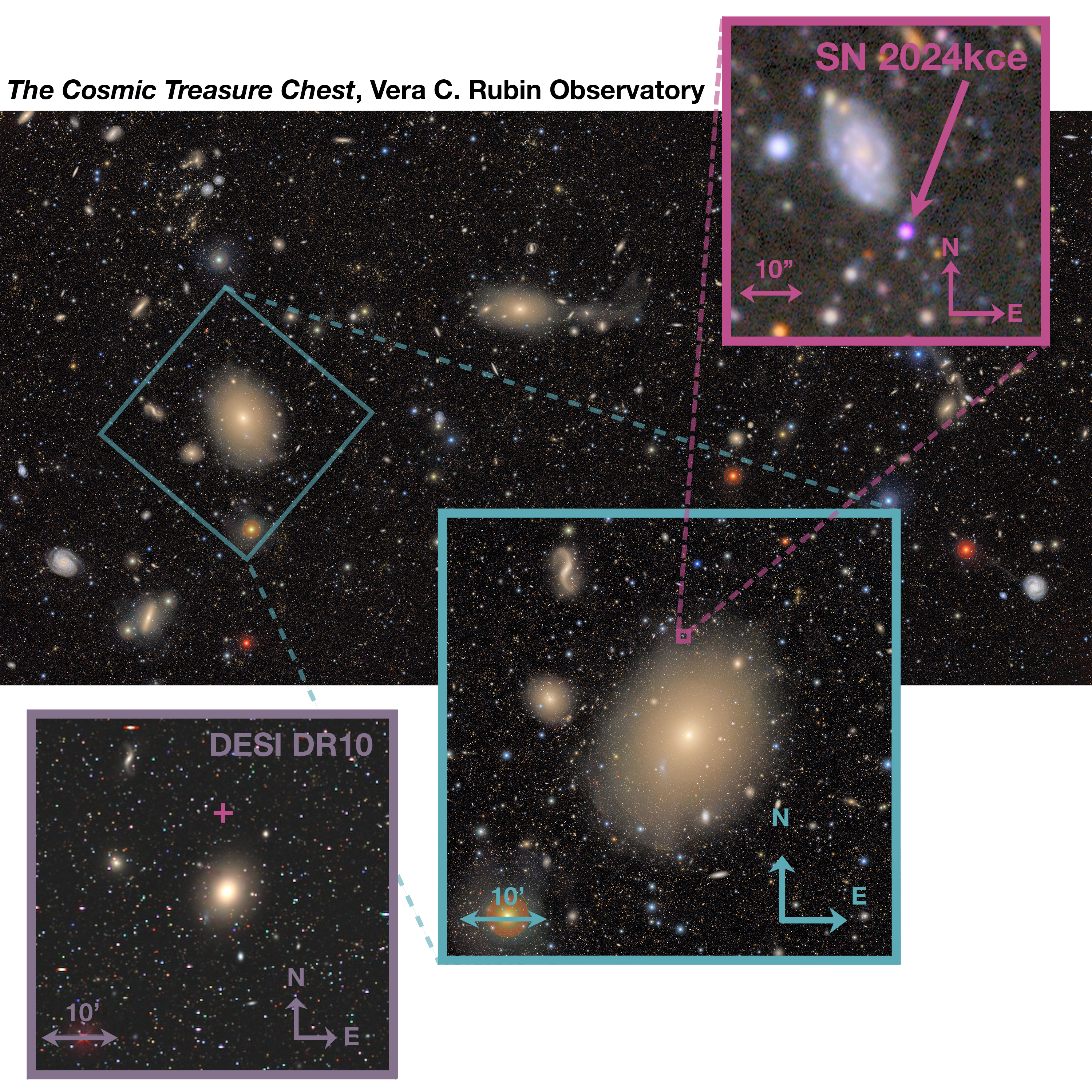}
    \caption{M49, the host galaxy of \kce. We examine the host M49 and location of \kce. The purple square shows the host imaged by DESI DR10 \cite{desi}; the location of \kce (pink plus), 61.4~kpc projected offset from galactic center, appears to be outside of the galactic halo. \kce was captured in the first look images for the Vera C. Rubin Observatory \cite{LSSTfirstlook} (the large panel). A zoom in of the host (teal square) shows that \kce resides within the diffuse stellar light of M49, in a remote stellar population with no evidence for an underlying satellite or other galactic substructure. A final zoom in (pink square) reveals the SN roughly a year after explosion. The spiral galaxy near \kce in the zoomed in region is a background galaxy, not associated with \kce or M49. Image adapted from The Cosmic Treasure Chest. NSF--DOE Vera C. Rubin Observatory/NOIRLab/SLAC/NSF/DOE/AURA
    }
    \label{fig:host}
\end{figure}
\afterpage{\clearpage}

\subsection*{\kce, a Peculiar Thermonuclear Transient:}

\begin{figure}[htbp]
\centering
    \includegraphics[width=\textwidth]{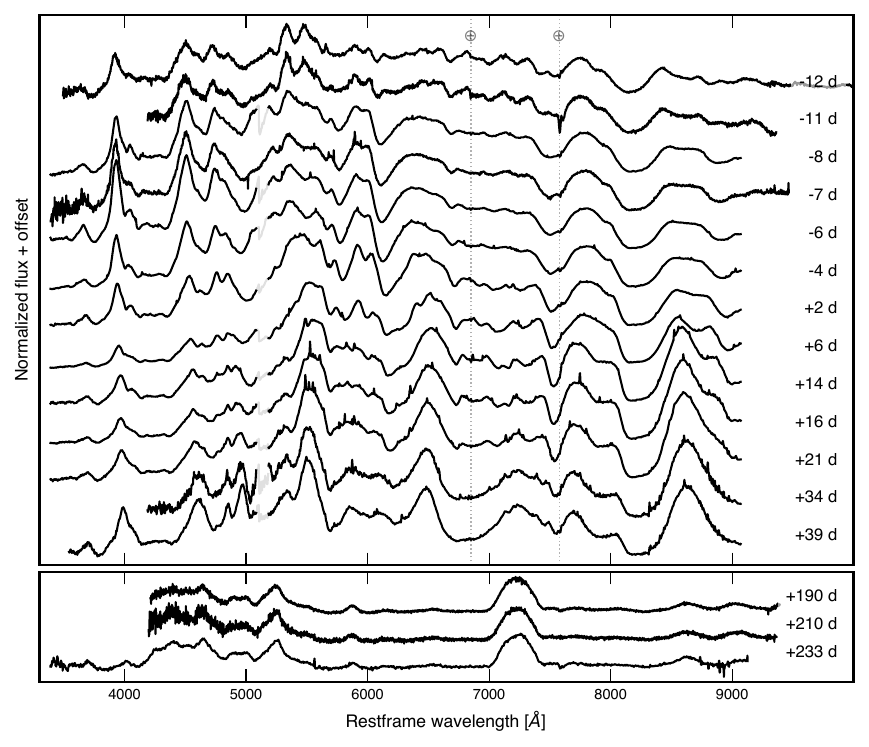}\vspace{-0.8em}
    \includegraphics[width=\textwidth]{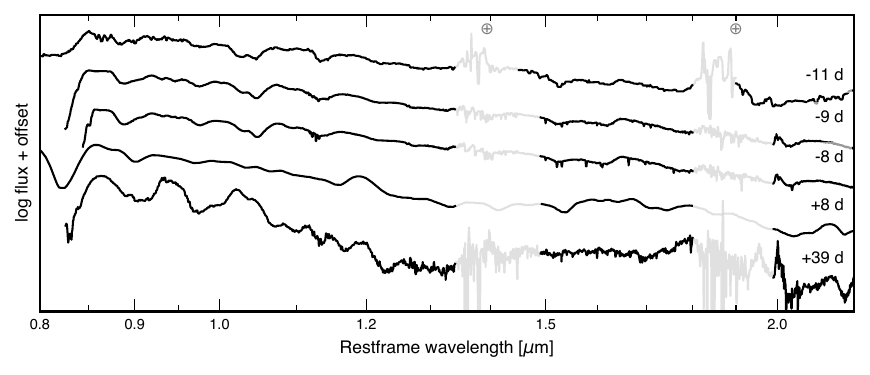}
\caption{Spectroscopic evolution of SN~2024kce. \textit{Top:} Optical spectra spanning $-12$ days before peak to $+233$ days after peak. Photospheric spectra are shown above and nebular spectra below. Greyed out flux denotes an artifact from the dichroic in the SNIFs spectra. Dashed lines show tellurics. \textit{Bottom:} NIR spectra spanning $-11$ days before peak to $+39$ days after peak. Greyed out regions denote un-recovered telluric contamination.
}\label{fig:obsspectra}
\end{figure}

\kce was discovered by the Zwicky Transient Facility (ZTF) on 02 June 2024 \cite{discoveryreport} in M49 (see Figure~\ref{fig:host}) and classified as a peculiar Type Ia SN
three days later \cite{classificationreport}. Including the classification spectrum, our spectroscopic follow-up began three days after discovery. The following night we obtained simultaneous optical and near-infrared (NIR) spectra eleven days before peak, while the outermost ejecta still shaped the spectrum. The series spans $-12$ to $+39$ days relative to peak, with three additional late-time spectra at $+190$, $+210$, and $+233$ days, taken after the SN entered the nebular phase (Figure~\ref{fig:obsspectra}). These are among the earliest panchromatic spectra obtained of a faint Type I SN and include the earliest NIR spectrum to date of a transient that evolves to show the strong nebular calcium emission characteristic of a calcium rich transient (Ca-RT).

Across its evolution, \kce does not map cleanly onto an established spectroscopic subtype. At early times it exhibits absorption from intermediate-mass elements including Si, S, and Ca, supporting a thermonuclear origin, together with the strong Ti absorption typical of 91bg-like SNe Ia. In the nebular phase, however, \kce develops the strong \fion{Ca}{2} emission seen in Ca-RTs. Weak \fion{Fe}{2} and \fion{Fe}{3} emission also remain present, indicating either a larger quantity of iron-group material, or that the iron-group material remains less mixed with the calcium, than in a typical Ca-RT \cite{2021ApJ...906...65P}.

The environment of \kce provides independent evidence for a thermonuclear origin. The transient lies 61.4~kpc in projection from the center of M49, a remote location similar to those of many Ca-RTs \cite{2014MNRAS.444.2157L}. In previous surveys, its position appeared to lie outside the galactic light entirely (purple square, Figure~\ref{fig:host}, left). Imaging obtained roughly one year after the explosion by the Vera C. Rubin Observatory shows that \kce instead lies within the diffuse stellar light of M49, with no visible underlying satellite (teal square, Figure~\ref{fig:host}, right). The small galaxy that sits near \kce in the most zoomed in panel of our figure (pink square, Figure~\ref{fig:host}) is a background galaxy with photo-$z=0.114\pm0.0336$ \cite{2016MNRAS.460.1371B, 2026arXiv260726149S} which we rule out as host. The remote stellar population at the location of \kce favors an old progenitor such as a WD, independently supporting the thermonuclear interpretation of its spectra.

\begin{figure}[htbp]
\centering
\includegraphics[width=\textwidth]{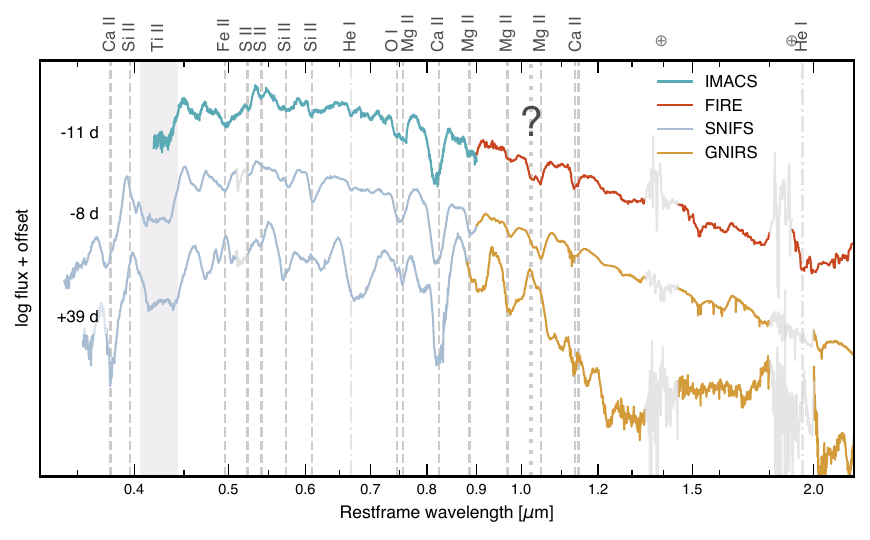}
\caption{Panchromatic optical and NIR spectra of \kce. Our panchromatic spectra consist of two early epochs (-11 and -8 days before peak) and one late time epoch (+39 days post peak). Common lines are drawn to guide the eye, blue-shifted to an ejecta velocity of 12,500 km/s (characteristic of the photosphere at -11 days). \ion{He}{1} features are shown at 16,400 km/s. We focus on the feature at $1.024~\mu$m (denoted with a question mark) and examine whether it arises from \ion{He}{1} or \ion{C}{1}.}
\label{fig:panchromatic}
\end{figure}

Our early panchromatic spectra provide a direct probe of the material in the outer ejecta. Figure~\ref{fig:panchromatic} shows the simultaneous optical and NIR spectra obtained $-11$ days before peak, together with two more panchromatic epoch at $-8$, and $+39$ days. Most absorption features in the earliest spectrum lie near the photospheric velocity of $12,500 \pm 170$~km/s  (measured from \ionl{Si}{2}{6355}~\AA), and are unusually narrow for such high velocities. In addition to the broad trough that sits around the photospheric velocity the Ca~II~NIR triplet also exhibits a high velocity component, by time of peak brightness this high velocity feature stabilizes at a velocity around 20,000~km/s becomes detached from the broader feature starting in our $+14$ day spectrum. The \ionl{Si}{2}{6355} feature also exhibits a similar feature at $17,000$~km/s, although here the high velocity component is detached from the earliest spectrum on. The NIR spectrum also contains a trough at $\lambda1.024~\mu$m that can be produced by either \ion{He}{1} or \ion{C}{1}. Distinguishing between these identifications requires confirmation from other spectral lines covered by the simultaneous optical and NIR spectra.

\subsection*{Helium in a Thermonuclear Transient:}
Identifying the $\lambda 1.024~\mu$m absorption requires distinguishing between two plausible contributors, \ionl{He}{1}{1.083~$\mu$m} and \ionl{C}{1}{1.069~$\mu$m}, the latter of which has been used to explain similar features in other thermonuclear SNe \cite{Hsiao2015, Pearson2024, 2021ApJ...914...57W, 2022ApJ...927..142L, Lu2023}. Our simultaneous optical and NIR spectra allow us to test these identifications through the other transitions of each ion.

\begin{figure}[hptb]
\centering
\includegraphics[width=\textwidth]{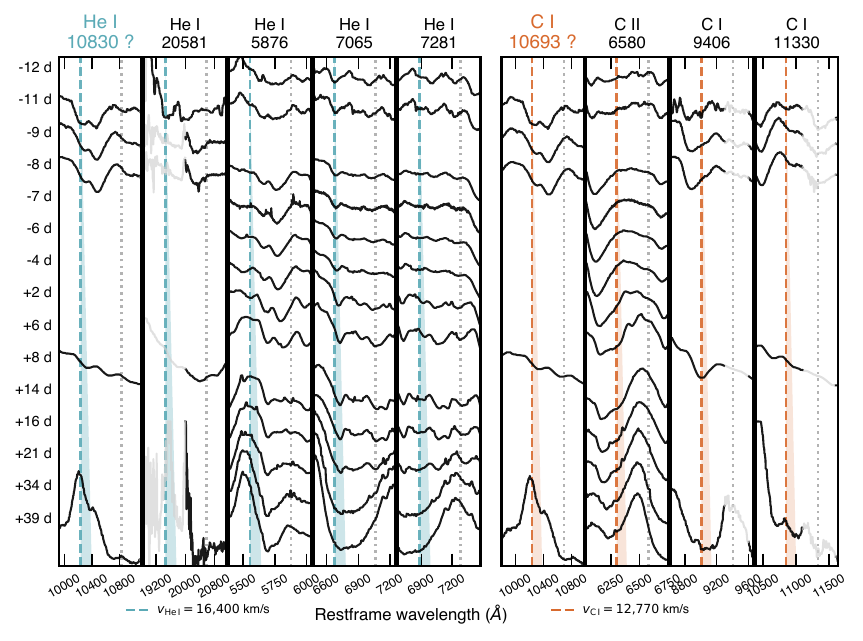}
\caption{Analysis of the $\lambda 1.024~\mu$m feature. The colored dashed lines represent the estimated velocities of \ion{He}{1} (blue) and \ion{C}{1} (orange) implied by the $\lambda$1.024 $\mu$m feature in our -11d NIR spectrum. The shaded wedges are placed to guide the eye to the feature minima along the right edge of the shaded region. The grey dotted lines show the rest frame wavelengths. Panels are plotted in order of expected line strength. We determine positive detections for \ion{He}{1} $\lambda$20581 and 7065~\AA, with tentative detections for \ion{He}{1} $\lambda$7281~\AA\ as well as \ion{C}{1} $\lambda$9406, and 11330~\AA. We therefore infer both \ionl{He}{1}{1.083}~$\mu$m and \ionl{C}{1}{1.069}~$\mu$m contribute to the feature seen at $\lambda$1.024~$\mu$m.   }
\label{fig:HeorC}
\end{figure}

In Figure~\ref{fig:HeorC}, we examine whether the $\lambda 1.024~\mu$m feature in our early NIR spectrum should be attributed to \ion{He}{1} or \ion{C}{1}. The two identifications imply different velocities in our earliest spectrum: a high-velocity \ion{He}{1} feature at 16,400~km/s or \ion{C}{1} near the photospheric velocity at 12,770~km/s. We trace the evolution of the feature through the optical and NIR spectra, searching for accompanying absorption at each of these velocities. The panels are ordered by expected line strength. We identify accompanying \ionl{He}{1}{2.058~$\mu$m} absorption in the NIR and \ionl{He}{1}{7065}~\AA\ in the optical at consistent velocities across the ions, with a tentative detection of \ionl{He}{1}{7281}~\AA. Together, these features support identifying the $\lambda 1.024~\mu$m trough with \ionl{He}{1}{1.083~$\mu$m}. We also note a tentative carbon detection. While there is no detection of \ionl{C}{2}{6580}~\AA\ which is typically the strongest optical carbon line in thermonuclear SNe, we do see hints of \ionl{C}{1}{9406}~\AA\ and \ionl{C}{1}{11330}~\AA, particularly in the earliest spectra, indicating that the carbon sits at a lower ionization level than typically seen in thermonuclear SNe. Therefore it is reasonable to conclude that carbon also contributes to the $\lambda 1.024~\mu$m feature.
The apparent contribution of both helium and carbon to the $\lambda 1.024~\mu$m NIR feature in \kce has important implications for interpreting similar features in other Type~I SNe. Identifying corroborating features from one element is not sufficient to rule out a contribution from the other.

In total, we conclusively identify \ion{He}{1} absorption at $\lambda 1.083~\mu$m and $\lambda 2.058~\mu$m in the NIR and \ionl{He}{1}{7065}~\AA\ in the optical, with a tentative detection of \ionl{He}{1}{7281}~\AA. This is the first thermonuclear supernova with helium absorption detected before maximum light, and the first in which the identification is verified by absorption at three separate wavelengths. The identification is further supported by the persistence of \ionl{He}{1}{1.083~$\mu$m} and \ionl{He}{1}{7065}~\AA\ across multiple pre-maximum epochs, allowing us to follow the helium as the ejecta evolves. The \ion{He}{1}~$\lambda 2.058~\mu$m NIR feature is particularly difficult to recover because it lies near a strong telluric band, requiring careful telluric correction. We recover it only with the observing strategy employed in the earliest NIR epoch, observed with the Folded-port InfraRed Echellette (FIRE) spectrometer \cite{2013PASP..125..270S}, emphasizing the importance of the early observations and careful telluric reduction for securing the identification.

In SNe Ib the strongest optical He feature is \ionl{He}{1}{5876}~\AA, but in \kce\ we do not identify a distinct absorption trough from this line. This region contains many blended lines, including \ionl{Si}{2}{5972}~\AA\ and the \ion{Ti}{2} forest, which could obscure the helium absorption. We see a shoulder at the expected wavelength, but without the other \ion{He}{1} lines present we would not identify it as a detection. The strengths of the optical \ion{He}{1} lines are not solely  proportional to the helium mass, depending instead on the distribution of radioactive material, the composition, and the radiation field \cite{2025ApJ...979..148L}. A weak \ionl{He}{1}{5876}~\AA\ therefore does not argue against the identification. \ionl{He}{1}{7065}~\AA, however, is present and grows in strength over time, as expected from the behavior of this line in SNe Ib \cite{2016ApJ...827...90L}.

The detection tells us something about where this helium must reside. Its velocity places it significantly above the photosphere, and the broad line with a lack of a P-Cygni feature indicates the helium lies within the SN ejecta rather than in a helium rich circumstellar medium. Therefore, the helium must lie in the outer most SN ejecta. \ion{He}{1} lines are not thermally populated at the temperatures found in the outskirts of SN ejecta; they require non-thermal electrons, produced when gamma rays from radioactive decay Compton scatter in the ejecta, to excite and ionize the helium \cite{1991ApJ...383..308L}. The helium we observe must therefore lie close enough to radioactive material to be excited by it, and only a modest helium mass mixed with that material is needed to produce a visible line \cite{2012MNRAS.422...70H}. In thermonuclear SNe the radioactive material is typically synthesized in the core of the explosion, not the outer layers. An exception is when the explosion is initiated by a helium-shell detonation, which leaves radioactive ash in the outermost ejecta. Together, the high velocity of the helium and its need for nearby radioactive material point to a detonated surface helium shell. 

\subsection*{The Nature of SN~2024kce:}
Typical C/O WDs retain little to no helium of their own, so this material must generally be supplied by a binary companion. In the double detonation scenario, the amount of helium on the surface of the C/O WD depends on the nature of that companion. Very little helium is transferred when the donor is a He WD or another C/O WD undergoing a dynamical merger (the D6 mechanism) \cite{Shen2018D6}. These thin shell systems can produce normal Type Ia SNe \cite{Shen2018,Polin2019} (see SN~2018aoz \cite{2022NatAs...6..568N, 2023ApJ...946....7N}). A non-degenerate companion, such as a He sub-dwarf, can donate substantially more helium before the explosion. These thick shell double detonations produce peculiar thermonuclear SNe with red colors at early times and can imprint strong line blanketing at blue wavelengths through peak brightness \cite{Polin2019, cattails} (see SN~2018byg \cite{De2019}). The shell mass required for detonation decreases as the WD mass increases \cite{Bildsten2007}. In addition to Type Ia SNe, a helium shell detonation has also been proposed for the Ia-like Ca-RT SN~2016hnk \cite{JacobsonGalan2020}, although a Chandrasekhar-mass origin has also been proposed \cite{Galbany2019}.

\begin{figure}[htbp]
\centering
\subcaptionbox{\kce (black) compared with our best fit double detonation model, a 0.69~$M_\odot$ WD with a 0.09~$M_\odot$ He shell. \textit{Top:} Optical spectra (teal). \textit{Bottom:} NIR spectra (orange) with a zoom-in on the $\lambda$~1.024$ \mu$m feature we attribute to \ionl{He}{1}{1.083}~$\mu$m. \label{fig:model-evolution}}{
    \includegraphics[width=\textwidth]{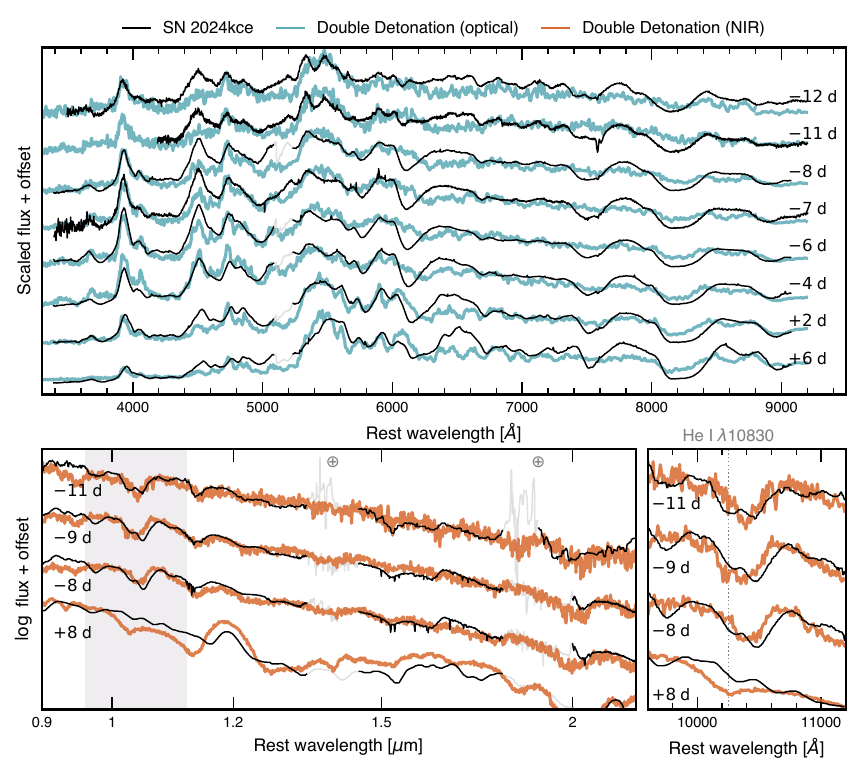}}
\subcaptionbox{The $\lambda$~1.024$ \mu$m feature modeled with four different combinations of NLTE species activated in the radiative transport calculations. The LTE version (purple) makes no feature in this region, and our NLTE helium (blue) and magneisum (green) runs show contributions from both elements in the $1 \mu$m region. The combination of both NLTE helium and magnesium (orange) reproduces the feature observed in \kce. \label{fig:model-NLTE}}{
    \includegraphics[width=\textwidth]{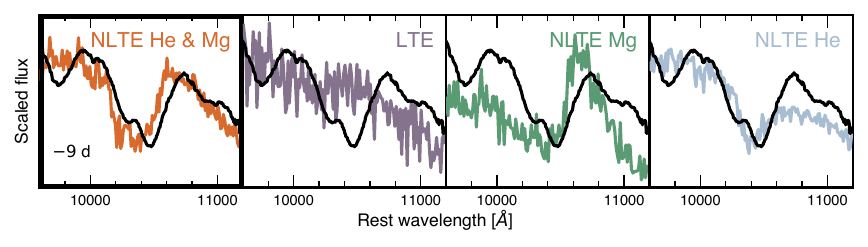}}
\caption{A double detonation model compared to the spectroscopic behavior of \kce.}\label{fig:model}
\end{figure}

With the exception of the late-time detection in SN~2022an \cite{2026ApJ..1004L..42C}, for which the lack of early data makes it difficult to establish a double detonation origin, no previous double detonation candidate had a conclusive detection of helium in its ejecta. This absence is not necessarily evidence that helium is missing: theoretical studies indicate that enough unburnt helium should remain in the outermost layers to produce a high velocity absorption feature \cite{2023MNRAS.524.4447C, 2025MNRAS.539.1404C}, and just as we see in \kce, such features are strongest in early NIR spectra, which have rarely been obtained.

The high velocity features \kce exhibits associated with Ca~II~NIR and \ionl{Si}{2}{6355} are also explained nicely by a double detonation. These features require a kinematically detached region of calcium and silicon which sit at in a higher velocity region than the bulk of the respective element. A double detonation can naturally produce this condition should the helium shell be large enough that intermediate mass elements are synthesized in the helium shell ashes that sit in the outermost ejecta.

To test a double detonation origin for \kce, we compare our observations with ejecta models computed using the moving-mesh hydrodynamics code \textsc{Cattails} \cite{2025ApJ...981...63H}. 
The model that best reproduces the spectroscopic evolution of \kce consists of a 0.69~$M_\odot$ C/O WD triggered by the ignition of a 0.09~$M_\odot$ helium shell. The explosion model retains 0.051~$M_\odot$ of unburnt helium between 13,300 and 22,000~km/s. This helium is co-located with the radioactive ash of the shell detonation, including $^{48}$Cr, $^{52}$Fe, and $^{56}$Ni. The model therefore places high velocity unburnt helium beside a source of non-thermal energy, as required by the line identification. The model also retains 0.01 M$_{\odot}$ unburnt carbon that lies in a thin layer just beneath the He shell ashes at a velocity of 12,300–13,300 km/s, consistent with the inferred velocity of the observed features. 

We compute the predicted spectra with the radiative transport code \textsc{Sedona} \cite{sedona}. Our treatment in \textsc{Sedona} is predominantly performed under the assumption of local thermodynamic equilibrium (LTE) with the exception of helium, and magnesium, for which we implement a full non-LTE (NLTE) treatment. This is vital for modeling our feature of interest as both \ion{He}{1}, and \ion{Mg}{2} require the non-thermal excitations not produced in LTE calculations. 

Figure~\ref{fig:model} shows the comparison between our \textsc{Sedona} results and \kce. The synthetic spectra match the optical evolution of \kce from $-11.8$ to $+6.2$ days, including the strong Ti and Fe absorption blueward of 5000~\AA~driving the red colors of \kce, which can result from line blanketing by helium burning ashes at blue wavelengths (Figure~\ref{fig:model-evolution}, \textit{top, teal}). The main systematic discrepancy is a lack of \ionl{O}{1}{7773}~\AA~in the optical spectrum, a feature known to depend sensitively on NLTE corrections \cite{2012MNRAS.424.2139D}. The strength of the \ionl{Si}{2}{6355} line is weaker in the model, but it does produce the high velocity component of the feature. For our three earliest epochs the model reproduces the NIR continuum as well as the absorption feature at $\lambda1.024~\mu$m (Fig~\ref{fig:model-evolution}, \textit{bottom, orange}). We start to see divergence between the model and \kce in the $+8$~d spectrum, likely due to the ejecta becoming optically thin in the NIR earlier than the optical, requiring more NLTE corrections to properly capture.

Figure~\ref{fig:model-NLTE} compares the $\lambda$1.024$ \mu$m feature in the -9 day \kce spectrum to our model when activating different choices of species in NLTE during the radiative transport calculations. We in turn try a full LTE calculation (purple), calculations where we turn on only NLTE magnesium and helium in turn (green and blue), and our final model where both helium and magnesium are treated in NLTE (orange). When the radiative transport is performed fully in LTE the $\lambda$1.024$ \mu$m feature is not created, so while there is carbon in the modeled ejecta, a contribution from \ionl{C}{1}{1.069}~$\mu$m is not strong enough to create this feature on its own. The runs in which magnesium and helium are each treated in NLTE alone show that both elements can contribute to the absorption trough. We note that these runs cannot be taken as additive, treating both elements in NLTE changes the electron density, temperature, and non-thermal energy budget that each species sees, so the combined spectrum is not the sum of the two separate NLTE models. Our final, combined, run shows that both NLTE magnesium and helium are necessary to achieve the feature as we see it in \kce. 

Throughout the photospheric evolution of \kce the double detonation model reproduces the red colors, the NIR continuum, the photospheric feature velocities, and the high velocity \ion{He}{1} absorption. The excellent agreement between these independent spectral constraints supports a helium shell double detonation origin for \kce.

\subsection*{Implications for Understanding the Thermonuclear Landscape:}

\begin{figure}[p]
\centering
\includegraphics[width=\textwidth]{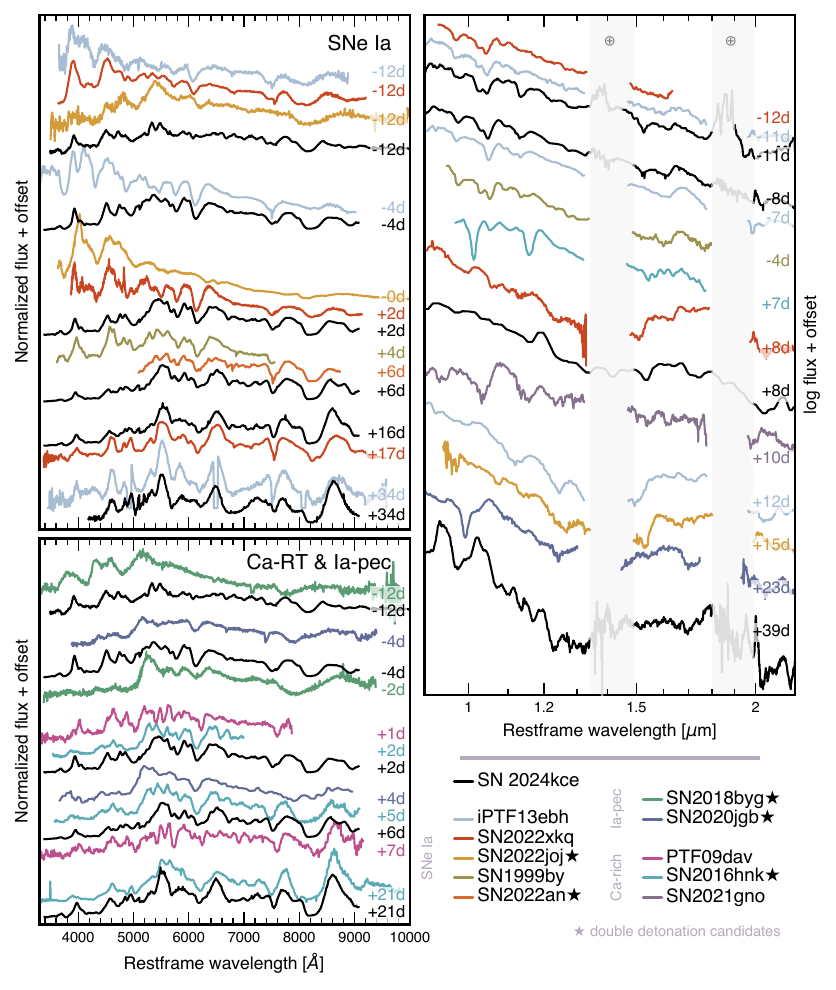}
\caption{SN~2024kce compared with fast declining SNe~Ia and Ca-RTs. Optical spectra are shown on the left (separated by SNe Ia on the top and Ca-RTs on the bottom) and NIR spectra on the right. \kce lies between the two classes, optically appearing more like a Ia-like Ca-RT, while resembling a sub-luminous SN Ia in the NIR. } \label{fig:otherSNe}
\end{figure}

Figure~\ref{fig:otherSNe} compares \kce with low luminosity, fast declining SNe Ia and Ca-RTs. No single comparison object resembles the full spectral evolution. At optical wavelengths, \kce exhibits the strong Ti and Fe absorption associated with 91bg-like SNe Ia. Its optical spectra also show a strong resemblance to the Ia-like Ca-RTs SN~2016hnk and PTF09dav. \kce, however, is considerably redder than the comparison SNe. The degree of blue suppression in \kce lies between these events and the thick he-shell double detonation candidate SN~2018byg, whose blue wavelengths are almost entirely blanketed through peak light \cite{De2019}. This intermediate amount of line blanketing is consistent with the helium shell ashes produced in the model presented above.

\afterpage{\clearpage}

The NIR spectra reveal a similarly mixed set of characteristics. The early NIR continuum shows a reasonable match to the SNe Ia, although absorption features are all stronger in \kce than other SNe Ia at similar phases. In the +8 day spectrum \kce exhibits a sharp rise in flux near $1.5~\mu$m, resembling an H-band break \cite{2019ApJ...878...86A}, although not nearly as strong as that seen in 91bg-like SN~2022xkq. This feature is consistent with blended Fe and Co absorption, providing an independent indication that \kce contains the iron-group ejecta characteristic of a thermonuclear explosion. At any given phase \kce looks less like the Ca-RTs in the NIR. Neither Ca-RT exhibits an H-band break, and SN~2016hnk shows overall stronger and narrower features than \kce. The Ca feature near near 1.2~$\mu$m is stronger in \kce than in the SNe Ia, but weaker than in the Ca-RTs. Perhaps most peculiar is the strong resemblance between the rising -8 day \kce spectrum and the post maximum +10 day spectrum of the Ib/c-like Ca-RT SN~2021gno. Several of the SNe Ia also show the double absorption feature near $\lambda 1.02~\mu$m that we attribute to \ion{He}{1} and \ion{Mg}{2} in \kce. In those events, the blue component has instead been attributed to \ion{C}{1} because no corroborating helium absorption was identified elsewhere in their spectra. 

SN~2022an provides the closest comparison to the helium absorption in \kce, although in SN~2022an the $\lambda$1.02$ \mu$m feature was first observed a month or more after maximum light at a velocity of $\sim13,000$~km/s \cite{2026ApJ..1004L..42C}. 
A direct velocity comparison remains difficult due to the lack in overlap of NIR spectral phases where the feature is strongest. We also note that \ionl{He}{1}{7065} becomes too blended to measure reliably in \kce beyond +21~days, while the absorption in SN~2022an persists to nearly +90~days. Perhaps indicating that there is more, or longer lived, radioactive material co-located with the helium in SN~2022an, hinting towards a larger helium shell. 

It has been suggested that the strong NIR feature at $\sim$1~$\mu$m in the $+23$~d spectrum of SN~2020jgb also arises from unburnt helium in a double detonation, which would require a velocity of $\sim$22,000~km/s \citep{2023ApJ...946...83L}. A similar feature is present in the $+7$~d spectrum of SN~2016hnk \citep{JacobsonGalan2020}. If this identification is correct, it would require the radioactive ash in the helium shell to extend to significantly higher velocities than in our model, since the helium must be co-located with the radioactive material that excites it. We note, however, that the red shoulder adjacent to this feature aligns much more closely with the helium velocities we infer for \kce. The Ca~II~NIR features in double detonation candidates SN~2020jgb and SN~2018byg span a considerably larger velocity range than in \kce, covering both high-velocity and photospheric components, and both were well fit by double-detonation models with more massive helium shells than we infer for \kce \citep{2023ApJ...946...83L,De2019}. It is therefore possible that the radioactive species in such massive shells extend to high enough velocities to excite helium where the SN~2020jgb feature forms. An NLTE investigation would be needed to establish whether a feature so qualitatively different from the one we model can arise by this mechanism.

The failure of any single comparison object to reproduce \kce is itself informative. Its Ti and Fe absorption and H-band break are characteristic of a subluminous thermonuclear SN, while its strong calcium and resemblance to SN~2016hnk place it among the Ia-like Ca-RTs. Because the helium shell double detonation mechanism has been proposed for both groups, identifying this explosion mechanism in an event with properties between them strengthens the case that low luminosity SNe Ia and at least some Ca-RTs share a common origin. The helium detection may extend this connection further. Ca-RTs are commonly separated into Ia-like and Ib/c-like events in part through the presence or absence of helium. The detection of helium in the outer ejecta of a thermonuclear explosion shows that helium absorption alone cannot distinguish between thermonuclear and stripped-envelope progenitors. Some events classified as Ib/c-like Ca-RTs may therefore also arise from thermonuclear explosions, although \kce does not establish that all Ca-RTs share a single origin.

The faint, fast transients that populate the edges of the thermonuclear population have been difficult to understand in part because so few have been observed. The Rubin Observatory will discover these events in far greater numbers \cite{2019ApJ...873..111I}, but \kce demonstrates that finding them is not sufficient. Reading the composition of their outermost ejecta requires optical and NIR spectra obtained together and early, while that material still shapes the spectrum. The feature near $\lambda1.02~\mu$m is ambiguous in isolation. Simultaneous optical and NIR spectroscopy reveals the corroborating \ion{He}{1} transitions needed to identify it as helium. The infrared is equally valuable at the opposite end of the evolution. Our nebular spectra already place \kce between the classes it resembles, with the strong \fion{Ca}{2} emission of a Ca-RT alongside iron-group emission that a typical Ca-RT alone would not produce. In the optical those iron features are heavily blended, and the abundances they imply are correspondingly uncertain. Late time nebular infrared spectroscopy could resolve this tension and for events like \kce it could establish whether the inner ejecta more closely resemble those of a subluminous SN Ia or something distinct. 
Looking forward, the combination of early panchromatic follow-up and late-time JWST spectroscopy is essential to bracket these explosions, constraining both the helium shell and the C/O core.

\section{Methods}\label{sec:methods}
\subsection{Observations}\label{subsec:obs}
SN~2024kce (ZTF24aaqaroi) was discovered by the Zwicky Transient Facility \citep{2019PASP..131a8002B,discoveryreport} on 2024 June 2 (MJD 60463.195) at $m_g = 17.55 \pm 0.05$ mag and classified as a peculiar Type~Ia supernova on 2024 June 5 \citep{classificationreport}. The last physically constraining non-detection was an ATLAS $o$-band limit of 19.51 mag at MJD 60462.340, followed by an ATLAS $c$-band detection at $18.22 \pm 0.16$~mag at MJD 60463.065 \citep{2018PASP..130f4505T}. These observations constrain the emergence of detectable emission to a 0.72~d interval. 

Phases throughout this work are quoted in rest-frame days relative to $r$-band maximum $m_r = 13.5$ at MJD 60478.076 (2024 June 17) shared by private communication with Sears et al. (in preparation).

\subsubsection{Optical spectroscopy}\label{subsubsec:optspec}
Our optical spectroscopic series consists of 16 spectra: 13 photospheric-phase spectra obtained from $-12$ to $+39$~d relative to $r$-band maximum and 3 nebular spectra obtained at $+190$, $+210$, and $+233$~d (Figure~\ref{fig:obsspectra} and Table~\ref{tab:speclog}). The first spectrum, taken on 2024 June 5 with the Kast double spectrograph on the Lick Shane 3~m telescope \citep{MillerStone1994}, was obtained from the TNS classification report \cite{classificationreport}. One photospheric-phase spectrum was obtained on 2024 June 6 with IMACS on the Magellan Baade 6.5~m telescope \citep{2011PASP..123..288D}. IMACS was also used to take the two nebular spectra on 2024 December 24 and 2025 January 13. Ten epochs were obtained with the SuperNova Integral Field Spectrograph (SNIFS) on the University of Hawai`i 2.2~m telescope \citep{2004SPIE.5249..146L} through the Spectroscopic Classification of Astronomical Transients (SCAT) survey \citep{2022PASP..134l4502T}. The $-7$~d spectrum was  observed using the Wide Field Spectrograph (WiFeS) which is mounted on the Australian National University (ANU)'s 2.3\,m telescope at Siding Springs Observatory \citep{2007Ap&SS.310..255D, 2010Ap&SS.327..245D, 2024PASA...41...57P}. Another nebular spectrum was obtained with the Keck Cosmic Web Imager (KCWI) \citep{2018ApJ...864...93M} on 2025 February 5.

\subsubsection{Near-infrared spectroscopy}\label{subsubsec:nirspec}
We obtained five NIR spectra between $-11$ and $+39$~d relative to $r$-band maximum. The earliest was obtained on UT 2024 June 6.12 ($-11.0$~d) with the Folded-port InfraRed Echellette (FIRE) \citep{2013PASP..125..270S} on the Magellan Baade 6.5~m telescope at Las Campanas Observatory. This observation began within an hour of the IMACS optical spectrum, providing the $-11$~d panchromatic view from 0.42 to 2.5~$\mu$m shown in Figure~\ref{fig:panchromatic}.

FIRE was configured in its high-throughput prism mode with the $0.6^{\prime\prime}$ slit, providing continuous coverage from 0.8 to 2.5~$\mu$m at resolving powers of $R\sim500$, $\sim450$, and $\sim300$ in the $J$, $H$, and $K$ bands, respectively. The target was nodded along the slit in an ABBA sequence, and the science frames were obtained in sample-up-the-ramp readout mode to suppress read noise.

Three additional spectra were obtained with GNIRS on Gemini North \citep{2006SPIE.6269E..4CE} on 2024 June 8.37, 9.32, and 26.26,  corresponding to phases of $-8.7$, $-7.8$~d, and $+39.2~d$ as part of the Hawai\okina i Infrared Supernova Study (HISS; \cite{2025ApJS..281...28M, 2025OJAp....8E.120H, 2025ApJ...988..209H}). The latter was obtained contemporaneously with a SNIFS spectrum, providing the $-8$~d panchromatic epoch shown in Figure~\ref{fig:panchromatic}. The final NIR spectrum was also obtained by HISS at $+8.2$~d with SpeX \citep{2003PASP..115..362R} on the NASA Infrared Telescope Facility.

\subsection{Data reduction}\label{subsec:reduction}
SNIFS data were reduced by the SCAT pipeline \cite{2022PASP..134l4502T}.
IMACS data were reduced with IRAF \citep{iraf1,iraf2}. Reductions included bias subtraction, flat fielding, wavelength calibration with arc lamp exposures obtained right before or after the SN observation, and flux
calibration with spectra of spectrophotometric standard stars observed during the same night as the science targets.  

SpeX data were reduced using the \textsc{spextool} \citep{2004PASP..116..362C} and \textsc{xtellcor} packages \citep{2003PASP..115..389V} following standard HISS reduction steps \cite{2025ApJS..281...28M, 2025OJAp....8E.120H, 2025ApJ...988..209H}.

The KCWI cubes were reduced with the Keck-supported Keck Cosmic Web Imager data reduction pipeline \citep{2023ascl.soft01019N}. The spectrum was extracted using a 0.8” radius aperture centered on the SN.

GNIRS data were reduced using standard procedures implemented in \textsc{PypeIt} \cite{pypeit,2020zndo...3743493P} following the standard HISS data reduction \citep{2025OJAp....8E.120H, 2025ApJ...988..209H}.
For the data taken on 2024 July 26, we use the calibration files from 2024 June 08. Features in the +39~day NIR spectrum should be viewed as approximate, and we limit the use of this spectrum to qualitative statements.

WiFeS data were reduced using the updated \textsc{pyWiFeS} reduction pipeline \citep{2014Ap&SS.349..617C,2024PASA...41...57P}, while flux calibration was obtained using a spectrophotometric standard star that was observed on the same night.

The FIRE observing strategy and reduction procedures are designed to recover features from the edges of the telluric region. This methodology was essential for recovering the \ionl{He}{1}{2.06}~$\mu$m feature in our $-11$~d NIR spectrum. Data were wavelength-calibrated using Ne and Ar arc-lamp exposures and reduced with the \textsc{firehose} IDL pipeline \citep{2013PASP..125..270S}. Telluric absorption was removed with the \textsc{xtellcor} package \citep{2003PASP..115..389V}. During the night we run an analysis to find the best choice for telluric standard at the time of our observation which selects for a A0V standard at as similar an airmass and sky location as the science object. We then observe that A0V standard using the same instrumental configuration immediately adjacent to the science exposures. Its intrinsic hydrogen Paschen and Brackett absorption was removed through comparison with a model spectrum of Vega, and the resulting telluric correction was applied to the SN spectrum. The standard also provided the flux calibration. This procedure was developed by previous studies \citep{2019PASP..131a4002H,2026ApJ..1000..178K}, for which the performance of the correction in the strong telluric region containing the blueshifted \ionl{He}{1}{2.058~$\mu$m} feature has been tested through the recovery of Paschen-$\alpha$ in SNe~II \citep{2019ApJ...887....4D}.

All spectra shown in this work were shifted to the host-galaxy rest frame using $z=0.003$. 

\subsection{Host association, distance, and extinction}\label{subsec:host}
\kce lies at $\alpha = 187.4565^\circ$ and $\delta = +8.213469^\circ$ (J2000), at a projected separation of 61.4~kpc from the center of M49 (NGC~4472), the brightest member of the Virgo Cluster. Vera C. Rubin Observatory First Look imaging shows that the SN position falls within the faint extended halo light of M49, with no obvious underlying satellite or galactic substructure (Fig.~\ref{fig:host}) \cite{LSSTfirstlook}. The exposure time, filters, and limiting surface brightness of the release image have not been reported, so we use it only qualitatively and do not derive a limiting luminosity for an undetected satellite.

A faint galaxy, SDSS J122950.02+081300.9, lies close to \kce in projection, and we considered whether it, rather than M49, could be the host. The SDSS DR20 catalog gives $z_{\rm phot} = 0.114 \pm 0.034$ for this object \citep{2026arXiv260726149S}, computed with a local linear regression method \citep{2016MNRAS.460.1371B}, placing it well behind the Virgo Cluster. Two independent checks rule it out as the host. First, the implied luminosity: at $z = 0.114$ the distance modulus is $\mu = 38.6$ mag,  
so the observed peak of $m_r = 13.5$~mag would correspond to $M_r \approx -25$, six magnitudes brighter than a normal SN~Ia ($M_r \approx -19.1$). At the M49 distance, by contrast, the peak corresponds to $M_r = -17.5$, $1.6$,mag fainter than a normal SN~Ia and typical of the fast-declining, subluminous Ia. Second, the ejecta kinematics would imply a photosphere velocity of ${\sim}43,700$~km/s ($0.15c$) (as measured from the \ionl{Si}{2}{6355} trough) as opposed to the more typical ${\sim}12,500$~km/s with host M49. Velocities of this magnitude are not seen in any thermonuclear or stripped-envelope supernova at these epochs. We therefore rule out this galaxy as the host.

Given the host identification of M49 we adopt a distance modulus of $\mu = 31.091 \pm 0.071$~mag \citep{2025ApJ...982...26A}, corresponding to a distance of $d = 16.5 \pm 0.5$~Mpc, and a heliocentric redshift of $z=0.003$. 

We apply no correction for host-galaxy extinction. The remote halo environment and the absence of Na~I~D absorption at the host redshift in our highest-S/N spectra support negligible host extinction. We therefore interpret the unusually red spectral shape of \kce as intrinsic to the SN rather than the result of dust reddening.

\subsection{Explosion models and radiative transfer}\label{subsec:models}

We simulate double detonation explosions of sub-Chandrasekhar mass white dwarfs using \textsc{Cattails} \citep{cattails}, a spherically symmetric moving-mesh nuclear hydrodynamics code \citep{2025ApJ...981...63H}. Nuclear energy generation is calculated with a 13-isotope reaction network spanning $^{4}$He to $^{56}$Ni, generated with \textsc{pynucastro} \citep{pynucastro2, pynucastro_development_team_2026_21934093}. 
In order to best approximate the multi-dimensional nature of a helium shell ignition in our hydrodynamic simulations we ignite the white dwarf at the most combustible region, the pre-mixed material at the transition from core to shell. Each model starts by flipping up to 20\% of the $^4$He, $^{12}$C, and $^{16}$O to $^{20}$Ne in a small region within the pre-mixed material and transferring the binding energy directly to thermal energy. The resulting hot spot will immediately burn past $^{20}$Ne to heavier intermediate mass elements and radioactive isotopes driving burning in the shell while sending a shockwave to the core to trigger the second detonation. In this manner we allow for the shell to burn and expand outwards for the full duration of shock wave convergence and core thermonuclear runaway, as seen in multi-D simulations (e.g. \cite{2021ApJ...919..126B, 2024A&A...683A..44M}). 

\begin{figure}[tpbh]
    \centering
    \includegraphics[width=\linewidth]{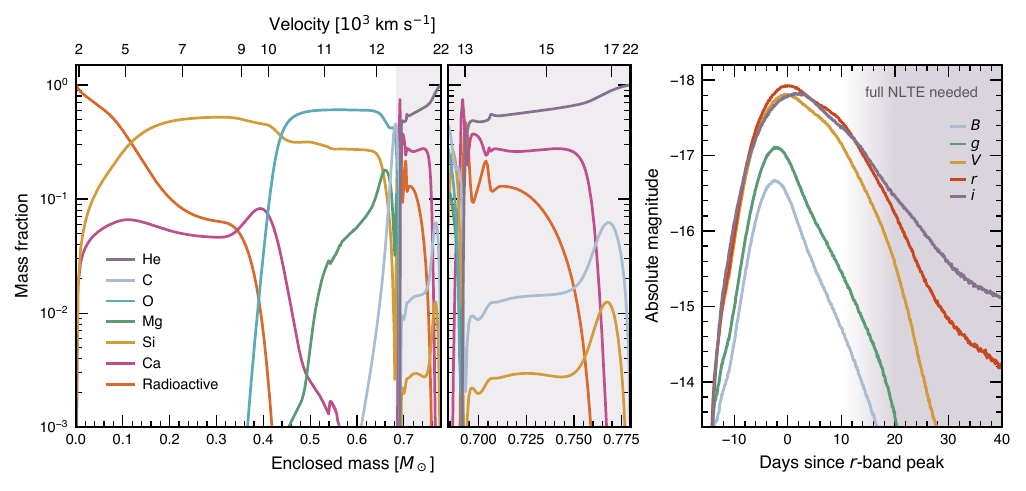}
    \caption{Additional aspects of our best fit double detonation model. \textit{Left:} Composition of the ejecta. The shaded region shown both in the left panel and in the zoomed in panel denotes the region above a  photosphere of $12,500$ km/s indicated by the velocities in the $-11$~d spectrum. This region begins just below the base of the helium shell ashes. \textit{Right:} \textsc{Sedona} light curves of our best fit model. The grey region denotes where the LTE assumption starts to break down as the ejecta begins to become optically thin. NLTE effects and emission features not produced by our LTE species contribute more to the ejecta flux as time from explosion increases. }
    \label{fig:extramodelstats}
\end{figure}

Our model grid spans total masses ($M_{WD} + M_{He}$) of 0.74 to 0.84 $M_\odot$ with helium shell masses of 0.05 to 0.14$M_\odot$.
Each calculation follows the helium shell detonation, shock convergence, and subsequent core detonation until the ejecta reach homologous expansion.

The homologous density, velocity, and composition profiles are mapped into the time dependent Monte Carlo radiative transfer code \textsc{Sedona} \citep{sedona} to calculate synthetic light curves and spectra. The calculations assume local thermodynamic equilibrium (LTE) for all elements except helium and magnesium, for which we solve the full non-LTE (NLTE) statistical equilibrium in every zone at every time step: level populations and ionization follow from radiative rates driven by the Monte Carlo radiation field, collisional rates with thermal electrons, and non-thermal ionization and excitation by the fast electrons from radioactive decay, while the remaining species take Saha-Boltzmann populations at the local temperature and electron density. The electron density and gas temperature are solved for the mixture as a whole, with the NLTE species entering the energy balance through their explicit heating and cooling rates and the LTE species through Kirchhoff's law. Atomic data are taken from the \textsc{cmfgen} compilation \citep{cmfgen}, including tabulated photoionization cross sections for helium. Because the electron density, temperature, radiation field, and non-thermal energy budget are shared by all species in a zone, the effects of treating helium or magnesium in NLTE are not additive, and we run the helium-only, magnesium-only, and combined cases as separate self-consistent calculations. We include energy contributions from the radioactive decay chains of $^{56}$Ni, $^{52}$Fe, and $^{48}$Cr. The latter two isotopes constitute most of the radioactive material synthesized in the helium shell and provide a local source of radioactive energy in the outer ejecta where the helium is distributed.

Among the calculated models, the model with a total mass of $0.78~M_\odot$, consisting of a $0.69~M_\odot$ C/O core and a $0.09~M_\odot$ helium shell, provides the closest agreement with the observed spectral evolution. The light curves and composition of this model are shown in Figure~\ref{fig:extramodelstats}. We selected this model by first comparing the spectral colors from continuum shape, the suppression of flux blueward of 5000~\AA, and the velocity of the principal optical absorption features from $-11.8$ to $+6.2$~d, then narrowing our selection using the peak $r$-band magnitude provided by Sears et al. (in prep). The final model is $0.2$~mag brighter at $r$-band maximum than \kce. Because the photometry of \kce is unpublished, the peak $r$-band magnitude is the only photometric constraint currently available, and the model was therefore not overly tuned to the light curve. Matching the observed peak would require fine tuning the helium shell mass at the $10^{-3}$~M$_{\odot}$ level on the basis of a single measurement. 

\subsection{Spectral line identification and velocity measurements}\label{subsec:velocities}
To measure the velocity evolution of the absorption complex near 1.02~$\mu$m, we normalize each NIR spectrum by a local linear pseudo-continuum anchored on either side of the feature. We fit the continuum-normalized spectrum from 9900 to 11100~\AA\ with two Gaussian absorption components using nonlinear least squares, masking regions affected by telluric absorption. 
We model the blue component as either \ionl{He}{1}{1.083~$\mu$m} or \ionl{C}{1}{1.069~$\mu$m} and identify the red component with \ionl{Mg}{2}{1.093~$\mu$m}. Under the helium identification, the blue component has a velocity of $16{,}400 \pm 120$~km/s at $-11.0$~d and declines to $14{,}100 \pm 290$~km/s by $+8.2$~d. Interpreting the same component as carbon would instead give $12{,}770 \pm 120$~km/s at $-11.0$~d, consistent with the photospheric velocity of $12{,}500 \pm 170$~km/s measured from \ionl{Si}{2}{6355} at $-10.9$~d. The \ion{Mg}{2} component has a velocity of $12{,}970 \pm 130$~km/s at $-10.9$~d, also consistent with the photosphere.

\subsection*{Data availability}

Upon publication the reduced optical and near-infrared spectra of SN 2024kce presented in this work will be made publicly available from the Weizmann Interactive Supernova Data Repository (WISeREP) \cite{Yaron2012} under the object name SN 2024kce. We will additionally include both observed and modeled data in a Zeondo deposit that will contain the Cattails ejecta profiles (density, velocity and composition) and the Sedona synthetic spectra and light curves for our model suite.

\subsection*{Code availability}
All data reduction pipelines used in this work are publicly available. 
\textsc{Cattails} is publicly available at \url{https://github.com/DHasenour/Cattails}. Input files will be made available in the Zenodo release upon paper acceptance for publication.

\begingroup
\setlength{\tabcolsep}{3pt}
\begin{table}[ht]
\caption{Spectroscopic observations of SN\,2024kce. Phases are in days relative to $r$-band maximum (MJD 60478.08).
Exposures are listed as $N_{\rm exp}\times t_{\rm exp}$ where the final
spectrum is a combination of individual frames. }
\label{tab:speclog}
\begin{tabular}{@{}llrlll l@{}}
\toprule
UT Date & MJD & Phase (d) & Telescope & Instrument & Range ($\mu$m) & Exp.\ (s) \\
\midrule
2024 Jun 5.23  & 60466.23 & $-11.8$ & Lick Shane 3\,m & Kast   & 0.35--1.03 &           \\
2024 Jun 6.12  & 60467.12 & $-11.0$ & Magellan Baade  & FIRE   & 0.80--2.50 & $8\times127$     \\
2024 Jun 6.16  & 60467.16 & $-10.9$ & Magellan Baade  & IMACS  & 0.42--0.94 & $2\times120$     \\
2024 Jun 8.37  & 60469.37 & $-8.7$  & Gemini-North    & GNIRS  & 0.83--2.53 & $4\times300$     \\
2024 Jun 9.32  & 60470.32 & $-7.8$  & Gemini-North    & GNIRS  & 0.85--2.53 & $4\times300$     \\
2024 Jun 9.36  & 60470.36 & $-7.7$  & UH 2.2\,m       & SNIFS  & 0.34--0.91 & 1800             \\
2024 Jun 10.40 & 60471.40 & $-6.7$  & ANU 2.3\,m      & WiFeS  & 0.34--0.95 & $600$            \\
2024 Jun 11.29 & 60472.29 & $-5.8$  & UH 2.2\,m       & SNIFS  & 0.34--0.91 & 1200             \\
2024 Jun 13.32 & 60474.32 & $-3.8$  & UH 2.2\,m       & SNIFS  & 0.34--0.91 & 1800             \\
2024 Jun 19.33 & 60480.33 & $+2.3$  & UH 2.2\,m       & SNIFS  & 0.34--0.91 & 1800             \\
2024 Jun 23.30 & 60484.30 & $+6.2$  & UH 2.2\,m       & SNIFS  & 0.34--0.91 & 1800             \\
2024 Jun 25.27 & 60486.27 & $+8.2$  & NASA IRTF       & SpeX   & 0.69--2.54 & $10\times120$    \\
2024 Jul 1.29  & 60492.29 & $+14.2$ & UH 2.2\,m       & SNIFS  & 0.34--0.91 & 1800             \\
2024 Jul 3.29  & 60494.29 & $+16.2$ & UH 2.2\,m       & SNIFS  & 0.34--0.91 & 1800             \\
2024 Jul 8.27  & 60499.27 & $+21.2$ & UH 2.2\,m       & SNIFS  & 0.34--0.91 & 1800             \\
2024 Jul 21.29 & 60512.29 & $+34.2$ & UH 2.2\,m       & SNIFS  & 0.34--0.91 & 1800             \\
2024 Jul 26.26 & 60517.26 & $+39.2$ & UH 2.2\,m       & SNIFS  & 0.34--0.91 & 1800             \\
2024 Jul 26.26 & 60517.26 & $+39.2$ & Gemini-North    & GNIRS  & 0.83--2.53 & $8\times200$     \\
2024 Dec 24.31 & 60668.31 & $+190.2$ & Magellan Baade & IMACS  & 0.42--0.94 & $2\times1800$    \\
2025 Jan 13.30 & 60688.30 & $+210.2$ & Magellan Baade & IMACS  & 0.42--0.94 & $3\times1200$    \\
2025 Feb 5.40  & 60711.40 & $+233.3$ & Keck II        & KCWI (blue)   & 0.34--0.56 & $2\times700+620$         \\
&&&&KCWI (red)& 0.56--0.92& $6\times300+65$\\
\botrule
\end{tabular}
The earliest optical spectrum is the TNS classification spectrum \cite{classificationreport}.
\end{table}
\endgroup

\begin{table}[ht]
\small
\setlength{\tabcolsep}{4pt}
\caption{Comparison spectra used in Figure~\ref{fig:otherSNe}. Phases are in days
relative to each object's maximum light, using the peak epochs adopted
in this work. Spectra were retrieved from WISeREP \citep{Yaron2012}
and the Transient Name Server.}
\label{tab:compspec}
\begin{tabular}{lll r@{\,}l l r}
\toprule
SN & Type & UT Date & \multicolumn{2}{c}{Phase (d)} & Telescope/Instrument & Ref. \\
\midrule
SN\,1999by  & Ia-91bg   & 1999 May 6  & $-4$   & (NIR) & MDM 2.4\,m/TIFKAM     & \cite{Hoeflich2002} \\
            &           & 1999 May 14 & $+4$   & (Opt) & FLWO 1.5\,m/FAST      & \cite{Matheson2008} \\
PTF09dav    & Ca-rich   & 2009 Aug 14 & $+1$   & (Opt) & WHT/ISIS              & \cite{Sullivan2011} \\
            &           & 2009 Aug 20 & $+7$   & (Opt) & P200/DBSP             & \cite{Sullivan2011} \\
iPTF13ebh   & SN Ia & 2013 Nov 15 & $-12$  & (Opt) & HCT/HFOSC             & \cite{Hsiao2015} \\
            &           & 2013 Nov 16 & $-11$  & (NIR) & Magellan Baade/FIRE   & \cite{Hsiao2015,Lu2023} \\
            &           & 2013 Nov 20 & $-7$   & (NIR) & Magellan Baade/FIRE   & \cite{Hsiao2015,Lu2023} \\
            &           & 2013 Nov 23 & $-4$   & (Opt) & NOT/ALFOSC            & \cite{Hsiao2015} \\            
            &           & 2013 Dec 9  & $+12$  & (NIR) & Magellan Baade/FIRE   & \cite{Hsiao2015,Lu2023} \\
            &           & 2013 Dec 31 & $+34$  & (Opt) & Magellan Baade/IMACS  & \cite{Hsiao2015} \\
SN\,2016hnk & Ca-Ia     & 2016 Oct 30 & $+2$   & (Opt) & SOAR/Goodman          & \cite{JacobsonGalan2020} \\
            &           & 2016 Nov 2  & $+5$   & (Opt) & NTT/EFOSC2            & \cite{Galbany2019} \\
            &           & 2016 Nov 4  & $+7$   & (NIR) & Gemini-S/FLAMINGOS-2  & \cite{Galbany2019} \\
            &           & 2016 Nov 18 & $+21$  & (Opt) & NTT/EFOSC2            & \cite{Galbany2019} \\
SN\,2018byg & Ia-pec    & 2018 May 8  & $-12$  & (Opt) & P200/DBSP             & \cite{De2019} \\
            &          & 2018 May 17 & $-2$   & (Opt) & P200/DBSP             & \cite{De2019} \\
SN\,2020jgb & Ia-pec       & 2020 May 13 & $-4$  & (Opt) & NOT/ALFOSC       & \cite{2023ApJ...946...83L} \\
            & (double-det) & 2020 May 21 & $+4$  & (Opt) & P60/SEDM         & \cite{2023ApJ...946...83L} \\
            &              & 2020 Jun 9  & $+23$ & (NIR) & Gemini-N/GNIRS   & \cite{2023ApJ...946...83L} \\
SN\,2021gno & Ca-rich   & 2021 Mar 29 & $+10$  & (NIR) & IRTF/SpeX             & \cite{JacobsonGalan2022} \\
SN\,2022an  & Ia-91bg   & 2022 Jan 7  & $+6$   & (Opt) & SOAR/Goodman          & \cite{JacobsonGalan2022tnscr,2026ApJ..1004L..42C} \\
SN\,2022joj & SN Ia        & 2022 May 11 & $-12$  & (Opt) & FTN/FLOYDS-N          & \cite{PadillaGonzalez2024} \\
            &           & 2022 May 23 & $+0$   & (Opt) & Lick Shane 3\,m/Kast  & \cite{Liu2023} \\
            &           & 2022 Jun 7  & $+15$  & (NIR) & Keck~II/NIRES         & \cite{Liu2023} \\
SN\,2022xkq & Ia-91bg   & 2022 Oct 17 & $-12$  & (Opt) & SOAR/Goodman          & \cite{Pearson2024} \\
            &           & 2022 Oct 17 & $-12$  & (NIR) & NTT/SofI              & \cite{Pearson2024} \\
            &           & 2022 Oct 31 & $+2$   & (Opt) & NOT/ALFOSC            & \cite{Pearson2024} \\
            &           & 2022 Nov 6  & $+8$   & (NIR) & Keck~II/NIRES         & \cite{Pearson2024} \\
            &           & 2022 Nov 13 & $+17$  & (Opt) & SALT/RSS              & \cite{Pearson2024} \\
\botrule
\end{tabular}
{The following SNe have been proposed as double detonation candidates: SN\,2016hnk \citep{JacobsonGalan2020}, SN\,2018byg \citep{De2019}, SN\,2020jgb \citep{2023ApJ...946...83L}, SN\,2022an \citep{2026ApJ..1004L..42C}, SN\,2022joj \citep{Liu2023,PadillaGonzalez2024}.}
\end{table}

\subsection*{Acknowledgments}

We thank Saurabh Jha for helpful discussions and for sharing the peak magnitude and time of maximum of \kce in advance of  publication.

This paper includes data gathered with the 6.5 meter Magellan Telescopes located at Las Campanas Observatory, Chile. This data was acquired from time awarded when A.P. was a postdoc supported by a Carnegie Fellowship.

Based in part on data acquired at the ANU 2.3-metre telescope under program 2410166. The automation of the telescope was made possible through an initial grant provided by the Centre of Gravitational Astrophysics and the Research School of Astronomy and Astrophysics at the Australian National University and through a grant provided by the Australian Research Council through LE230100063. The Lens proposal system is maintained by the AAO Research Data \& Software team as part of the Data Central Science Platform. We acknowledge the traditional custodians of the land on which the telescope stands, the Gamilaraay people, and pay our respects to elders past and present.

Some of the data presented herein were obtained at Keck Observatory, which is a private 501(c)3 non-profit organization operated as a scientific partnership among the California Institute of Technology, the University of California, and the National Aeronautics and Space Administration. The Observatory was made possible by the generous financial support of the W. M. Keck Foundation. The raw Keck Observatory data presented herein, once public, can be accessed from the Keck Observatory Archive (KOA), which is operated by the W. M. Keck Observatory and the NASA Exoplanet Science Institute (NExScI), under contract with the National Aeronautics and Space Administration. The authors wish to recognize and acknowledge the very significant cultural role and reverence that the summit of Maunakea has always had within the Native Hawai\okina ian community. We are most fortunate to have the opportunity to conduct observations from this mountain.

\subsection*{Funding Acknowledgments}

A.P. acknowledges support from JWST-AR-10953.001-A awarded by the Space Telescope Science Institute (STScI), which is operated by the Association of Universities for Research in Astronomy, Inc., for NASA, under contract NAS5-03127.
J.H. acknowledges support from the NASA Hubble Fellowship program.
D.O.J. acknowledges support from NSF grants AST-2407632, AST-2429450, AST-2510993, and AST-2607081, NASA grants 80NSSC24M0023 and 80NSSC24K0353, and JWST-GO-05324.031, awarded by the Space Telescope Science Institute (STScI), which is operated by the Association of Universities for Research in Astronomy, Inc., for NASA, under contract NAS5-26555. This work is also funded in part by the Gordon and Betty Moore Foundation through Grant GBMF13900 to D.O.J.

\bibliography{bibliography}

\end{document}